\documentclass[journal=cmatex,manuscript=article,layout=traditional]{achemso}
\usepackage{graphicx}
\usepackage{dcolumn}
\usepackage{bm}
\usepackage{times}
\usepackage{epstopdf}
\usepackage[version=3]{mhchem} 
\usepackage{xcolor}

\usepackage{mathrsfs}
\usepackage{lipsum}
\usepackage{amsmath}
\usepackage{amsfonts}
\usepackage{array}
\usepackage{color}
\usepackage{ulem}

\newcommand{\degC}{\,$^{\circ}$C}

\title{Enhanced N\'{e}el temperature and unusual thermal expansion in flux-grown FeCrAs crystals
\footnote{Notice: This manuscript has been authored by UT-Battelle, LLC, under contract DE-AC05-00OR22725 with the US Department of Energy (DOE). The US government retains and the publisher, by accepting the article for publication, acknowledges that the US government retains a nonexclusive, paid-up, irrevocable, worldwide license to publish or reproduce the published form of this manuscript, or allow others to do so, for US government purposes. DOE will provide public access to these results of federally sponsored research in accordance with the DOE Public Access Plan (https://www.energy.gov/doe-public-access-plan).}
}

\author{Michael A. McGuire}
\email{mcguirema@ornl.gov}

\author{Matthew S. Cook}
\affiliation{Materials Science and Technology Division, Oak Ridge National Laboratory, Oak Ridge, Tennessee 37831 USA}

\author{Brenden R. Ortiz}
\affiliation{Materials Science and Technology Division, Oak Ridge National Laboratory, Oak Ridge, Tennessee 37831 USA}

\author{Jiaqiang Yan}
\affiliation{Materials Science and Technology Division, Oak Ridge National Laboratory, Oak Ridge, Tennessee 37831 USA}

\author{Andrew F. May}
\affiliation{Materials Science and Technology Division, Oak Ridge National Laboratory, Oak Ridge, Tennessee 37831 USA}


\begin{document}


\begin{abstract}
We report results from our experimental investigation of the distorted-kagome compound FeCrAs. For this work, we used tin metal as a flux to produce needlelike crystals, which we characterized using single crystal x-ray diffraction as well as measurements of magnetization, electrical transport, and heat capacity. The physical behaviors differ in two notable ways from those of previously studied crystals grown from a stoichiometric melt. First, the N\'{e}el temperature is found to be 150\,K, about 25\,K higher than in previous reports. Second, the Sommerfeld coefficient, a measure of the electronic heat capacity, is found to be about half of the previously reported value. These differences indicate stronger magnetic interactions and fewer charge carriers in the flux-grown crystals, which may be related to differences in stoichiometry or disorder. In addition, we find unusual thermal expansion behavior, with an anomaly at the N\'{e}el temperature and nearly temperature independent thermal expansion along the hexagonal \textit{c}-axis above this transition. This suggests significant spin-lattice coupling, which may provide insight into non-metallic transport properties that have been associated with anomalous charge carrier scattering. The flux growth presented here may provide a useful approach to exploring the relationships between crystal chemistry and magnetism, transport, and spin-lattice coupling in this interesting material.

\end{abstract}


\section{Introduction}

The compound FeCrAs is a member of a large family of materials that adopt the ZrNiAl structure type. This non-centrosymmetric, hexagonal structure (space group $P\overline{6}2m$) displays extreme chemical versatility; reported ZrNiAl-type compounds include all transition metals except Tc, all lanthanides except Pr, and all group 3A, 4A, and 5A elements except C and N \cite{Villars-2024}. It is among the most common structure types adopted by intermetallic compounds \cite{Dshemuchadse-2015}. The ZrNiAl structure is an ordered variant of the \ce{Fe2P} structure. It is commonly formed with pnictogens, and there are at least 60 ZrNiAl-type ternary compounds with P, As, Sb, or Bi \cite{Villars-2024}. Treating the pnictogens as anions, such compounds contain two metal sites distinguished by their coordination: square pyramidal and tetrahedral. It has been noted that is P and As containing phases the more electropositive element prefers the site with square pyramidal coordination \cite{Fruchart-1982}. Examples of polymorphism and competition with related 1:1:1 structure types add to the chemical interest in these materials \cite{Kussmann-1998, Oliynyk-2025, Bao-2021, Kong-2024, Roy-Montreuil-1972}.

Notably, the ZrNiAl structure contains a distorted kagome net, occupied by Zr in the structure type's namesake. The structure can be derived from the prototypical kagome metal CoSn \cite{Meier-2020}. CoSn contains a kagome net of Co with Sn atoms in the centers of the Co hexagons, and these layers are separated by a honecomb layer of Sn. The ZrNiAl structure type is realized by stuffing triangular units into the hexagonal holes in this honeycomb layer. This distorts the neighboring kagome layers, producing the structure shown in Figure \ref{fig:structure}. This twisting-like distortion breaks the inversion symmetry of the crystal structure. A similar distortion has been found in the CoSn analogue RhPb, and the resulting noncentrosymmetric structure produces chiral phonons, topological phonons, and associated surface states \cite{Ptok-2021, Ptok-2023}.

Due in part to the chemical adaptability noted above, the ZrNiAl structure type supports a wide variety of interesting physical phenomena. Examples includes superconductivity \cite{Barz-1980, Inohara-2016, Kumar-2022, Su-2021}, Kondo physics \cite{Xie-2022}, and nontrivial band topology \cite{Laha-2020}. FeCrAs, the topic of the present study, is an example of a ZrNiAl-type compound with itinerant magnetism and unusual electronic properties.

The interesting behaviors of FeCrAs were first identified experimentally by Wu \textit{et al}., who developed a procedure for extracting single crystals from a stoichiometric melt (the compound does not melt congruently) and characterized them using electrical transport, heat capacity, magnetization, and neutron diffraction measurements \cite{Wu-2009, Swainson-2010, Wu-2011}. Those studies identified an antiferromagnetic phase transition and interesting electronic behavior. The electrical resistivity was shown over a very large temperature range, from 800\,K down to 100\,mK, to be ``non-metallic'' (that is, neither activated nor metallic, but increasing on cooling with a power law like temperature dependence). The low temperature electronic heat capacity was noted to be linear in temperature, as expected for a normal metal. The magnetic order in this compound is also interesting. It has been shown that FeCrAs has a noncollinear magnetic structure, with small ordered magnetic moments on the Cr atoms and no ordered moment detected on Fe by neutron diffraction or M\"{o}ssbauer spectroscopy \cite{Wu-2009, Swainson-2010, Jin-2019, Huddart-2019, rancourt1984hyperfine}. This interesting observation had been previously predicted based on electronic structure calculations \cite{Ishida-1996}. Scattering related to magnetic fluctuations has been proposed as a possible explanation for the unusual behavior seen in resistivity \cite{Rau-2011, Akrap-2014, Plumb-2018}. Most recently, FeCrAs was identified as a candidate material to realize altermagnetism in a non-collinear spin structure \cite{Singh-2025}.

Inspired by these interesting transport and magnetic behaviors observed in FeCrAs grown from a stoichiometric melt, we have investigated crystals grown by a complementary growth technique, a molten metal flux \cite{Kanatzidis-2005}. Flux growths can enable production of high-quality crystals of materials that are otherwise hard to obtain, and especially those that decompose incongruently. For the present work, needlelike crystals were grown from a Sn flux and characterized using electrical transport, magnetization, and heat capacity measurements, as well as single crystal x-ray diffraction. We find the magnetic ordering temperature, which occurs at 150\,K in the flux grown crystals, is about 25\,K higher than in previous reports, which range from 100 to 125\,K \cite{Wu-2009, Wu-2011, Plumb-2018, Huddart-2019, Jin-2019}. The electronic heat capacity differs as well. In addition, we observed an anomaly in the thermal expansion at the N\'{e}el temperature as well as an unusually weak temperature dependence in the \textit{c}-axis lattice constant above the transition. This provides evidence of spin-lattice coupling in FeCrAs that may provide new insights into the unusual electronic behavior of this interesting material.

\section{Crystal growth and structure}

\begin{table}
\caption{Crystal structure refinement results for FeCrAs from single crystal x-ray diffraction. The space group is $P\overline{6}2m$. Fe is at Wyckoff position 3g ($x_{\rm Fe}$, 0, $\frac{1}{2}$), Cr is at 3f ($x_{\rm Cr}$, $x_{\rm Cr}$, 0), As1 is at 2d ($\frac{2}{3}$, $\frac{1}{3}$, $\frac{1}{2}$), As2 is at 1a (0, 0, 0).
}
\begin{tabular}{lcc}
\hline															
\textit{T} (K)	&	201(2)	&	101(2)	\\
\textit{a} ({\AA})	&	6.08890(10)	&	6.08200(10)	\\
\textit{c} ({\AA})	&	3.65840(10)	&	3.65630(10)	\\
\textit{V}  ({\AA}$^3$)	&	117.463(5)	&	117.129(5)	\\
$x_{\rm Cr}$	&	0.5846(2)	&	0.5851(1)	\\
$x_{\rm Fe}$	&	0.7522(1)	&	0.7523(1)	\\
Data / parameters	&	428 / 15	&	427 / 15	\\
R1 (all data)	&	0.0271	&	0.0248	\\
wR2 (all data)	&	0.0598	&	0.0560	\\
\hline															
\end{tabular}\
\label{Table}
\end{table}

\begin{figure*}
\begin{center}
\includegraphics[width=6.65in]{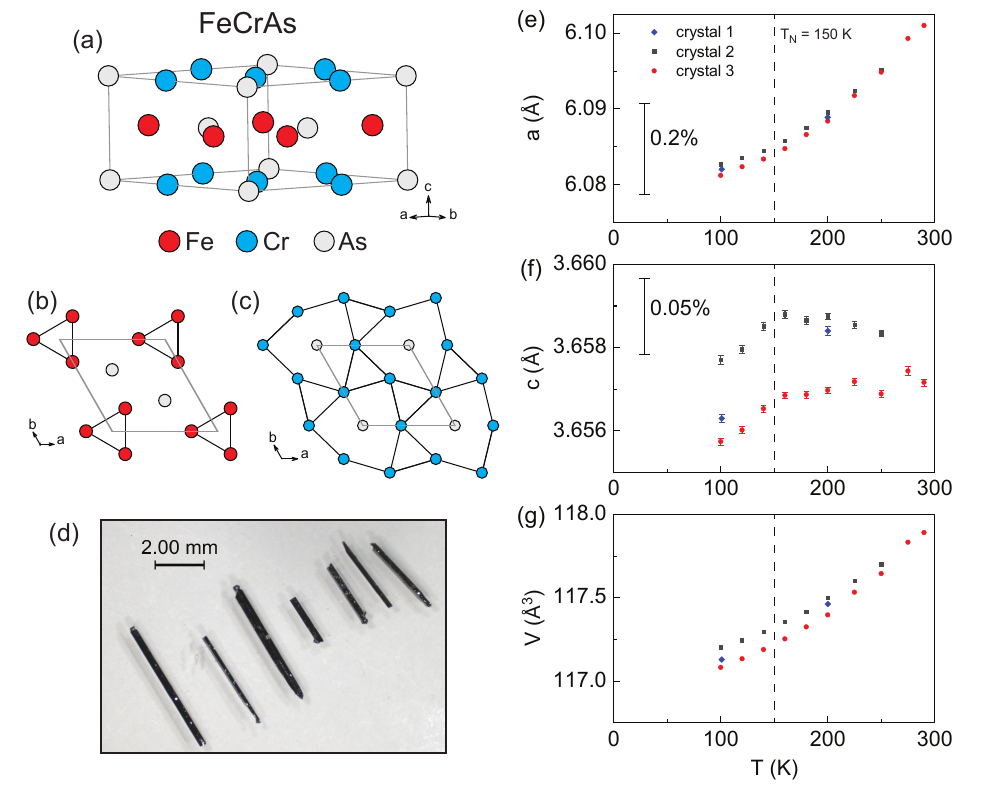}
\caption{\label{fig:structure}
The crystal structure of FeCrAs. (a) The full unit cell. (b) The triangular arrangement of Fe atoms in the \textit{z} = 1/2 plane. (c) The distorted kagome net of Cr atoms in the \textit{z} = 0 plane. (d) A selection of FeCrAs crystals grown from a Sn flux. (e-g) Lattice parameters and unit cell volume determined from single crystal x-ray diffraction on three FeCrAs crystals. Error bars on data in panels (e) and (g) are smaller than the data markers. Dashed vertical lines mark the antiferromagnetic ordering temperature for the crystals.
}
\end{center}
\end{figure*}

To date, the crystals used for studying FeCrAs have been produced using an approach inspired by the growth of \ce{Fe2As} reported by Katsuraki and Achiwa \cite{Katsuraki-1966}. Wu \textit{et al}. refined the procedure for FeCrAs \cite{Wu-2009, Wu-2011}. They found that the material melts incongruently, but identified conditions that gave sizable FeCrAs crystals. The crystals produced this way are embedded an inhomogenous boule and are relatively easily separated from the other phases.

For the present work we grew FeCrAs using Sn as a flux. Iron granules (99.98\%), chromium powder (99.97\%), arsenic pieces (99.99999\%), and tin shot (99.99\%) were combined in a 1:1:1:50 molar ratio in alumina frit-disc crucible sets \cite{Canfield-2016} and sealed in silica ampoules under vacuum for standard flux growth \cite{Kanatzidis-2005}. The ampoules were heated to 1100\degC\ over 12 h, held at that temperature for 48 h, then cooled to 700\degC\ at 3.3\degC\ per hour (over 120 h). At 700\degC\ the ampoules were removed from the furnace and inverted into a centrifuge to separate the Sn-rich liquid flux from the FeCrAs crystals. This technique produced numerous needle shaped crystals of FeCrAs as shown in Figure \ref{fig:structure}d. Small amounts of flux seen on some of the crystal surfaces could be removed by soaking in a 1:1 solution of HCl and water. As discussed below, some of the crystals also contained inclusions in the form of Sn-rich cores.

Variations of this procedure were explored to a limited extent. We performed growths in which the temperatures were the same as above but the cooling rate was increased to 6.6 \degC\ per hour. These faster-cooled growths were done with a pure Sn flux, with a flux containing 10\% Pb by weight, and with a flux containing 10\% Sb by weight.

Energy dispersive spectroscopy measurements using a scanning electron microscope (SEM) showed the crystals to have close to an equimolar composition, but a clear excess of Cr relative to Fe (Fe:Cr:As = 30:36:34 in atomic percent). This composition was the same for crystals from all four growth conditions used, and we observed no significant variation from crystal to crystal. To test the reliability of this quantification, similar measurements were conducted over a mm-scale area on a polished surface of a small arc-melted sample of composition FeCr. That measurement yielded a composition of 50:50, indicating that the FeCrAs crystals indeed contain an excess of Cr. Thus, we conclude that the composition of the crystals is close to Fe$_{0.9}$Cr$_{1.1}$As. No evidence of incorporation of Sn, Sb, or Pb from the fluxes into the FeCrAs phase was seen.

An early report and characterization of this phase was given by Hollan in a study of (Fe$_{\rm 1-x}$Cr$_{\rm x}$)$_{\rm 2}$As \cite{Hollan-1966}. While the endmembers of this series \ce{Fe2As} and \ce{Cr2As} adopt the tetragonal \ce{Cu2Sb} structure type, a hexagonal structure was found in the region $0.3 < x < 0.7$ corresponding to the ZrNiAl structure type. Thus, some compositional variability may be expected in the FeCrAs phase, and the flux growths reported here may prove convenient for studies of crystals with varying Fe:Cr ratios. With elemental Sn, the primary flux used here, Fe has a significantly higher solubility than Cr in the 700$-$1100\degC\ temperature range. This may lead to variation the precipitation rates for the two metals that would be consistent with the observed crystal composition.

The crystal structure was refined from x-ray diffraction data collected near 100 and 200 K. The previously reported structure was confirmed for our flux grown crystals. Results are collected in Table \ref{Table} and in crystallographic information files included as Supporting Information. No significant structural change, other than thermal expansion, was observed between these two temperatures. Analysis of powder neutron diffraction data in previous reports has shown that FeCrAs is stoichiometric and the Fe and Cr are well ordered on to the 3g and 3f sites, respectively \cite{Wu-2009, Swainson-2010}. Since the contrast between Cr and Fe is weak for x-ray diffraction, we cannot refine the Fe and Cr occupations freely; however, our data is consistent with the expected overall atomic siting as listed in the table. Using the data collected at 100 K, a model with the Fe and Cr sites swapped gave R1 = 0.0309 and wR2 = 0.0791, significantly worse than the agreement factors shown in Table \ref{Table}. Note the composition was left fixed at 1:1:1 for all refinements; any small excess of Cr could not be reliably modeled.

The structure is shown in Figure \ref{fig:structure} and has been described in detail in previous publications \cite{Hollan-1966, Swainson-2010, Wu-2011}. Briefly, the iron atoms are arranged in triangular trimers at the corners of the unit cell (Fig. \ref{fig:structure}c), and the Cr atoms form a distorted kagome net in the \textit{z} = 1/2 plane (Fig. \ref{fig:structure}d). Considering the As atoms as anions, the Cr and Fe sites are seen to be in square-pyramidal and tetrahedral coordination, respectively.

While there is no change in symmetry detected between 100 and 200\,K (across the magnetic ordering transition), unusual thermal expansion behavior is observed. Figure \ref{fig:structure}(e-g) show results from unit cell refinements from single crystal diffraction measurements conducted between 100 and 290\,K. Results are shown for two crystals selected from two separate growths (crystals 2 and 3) as well as crystal 1 that was used to collect data for Table \ref{Table}. The \textit{a} lattice parameter has normal thermal expansion behavior, changing by about 0.3\%. For reference, this is similar to the change seen in copper over this temperature range. However, a change in slope is seen near 150\,K. Notably, this is the N\'{e}el temperature in these crystals, as demonstrated in the physical properties discussed below. More striking behavior is seen in the \textit{c}-axis length (Fig. \ref{fig:structure}f), which is nearly temperature independent above 150\,K. Below this temperature it shows normal thermal expansion.
The unit cell volume V (Fig. \ref{fig:structure}g) varies smoothly with temperature over this range with no discernable features.

This change in thermal expansion behavior upon crossing $T_N$ indicates coupling between the magnetism and crystal lattice in FeCrAs. Previous optical measurements of infrared active vibrational modes also revealed anomalous temperature dependence, which the authors identified as evidence of spin-phonon or spin-lattice coupling \cite{Akrap-2014}. It would be interesting to examine this further to understand how it may be related to some of the unusual behaviors that have been noted in this compound. The weak temperature dependence of \textit{c} is also remarkable and motivates further study of the thermal expansion in this material over a wider temperature range. Perhaps it is related to magnetic correlations near and above $T_N$, which may be related to the unusual temperature dependence of the resistivity at high temperature in this ``non-metallic metal'' \cite{Wu-2009,Wu-2011,Akrap-2014}. It is worth noting that negative or unusual thermal expansion behavior is well known in magnetic materials \cite{Song2021}, and interesting lattice behaviors related to magneto-elastic coupling have been observed in compounds chemically and structurally related to FeCrAs \cite{Zach1990, Hu2019}. Such behaviors can be attributed to magnetovolume effects that couple bond distances to magnetic order/disorder \cite{Song2021}. The frustrated antiferromagnets \ce{Mn3Ge} and \ce{Mn3Sn} with kagome nets of Mn show anomalous thermal expansion near and below their ordering temperatures \cite{Song2018}. CrAs is another example of an antiferromagnet that experience anomalous thermal expansion associated with the development of magnetic order due to coupling of spin, lattice, and electronic degrees of freedom \cite{Hu2019}. In these cases, the main effect observed is a negative thermal expansion over a temperature range below the magnetic ordering temperature. In contrast, for FeCrAs, it appears that the more unusual behavior occurs above the magnetic phase transition, again motivating more detailed crystallographic studies of this material over a wider temperature range.

\section{Physical properties}

\begin{figure}
\begin{center}
\includegraphics[width=3.25in]{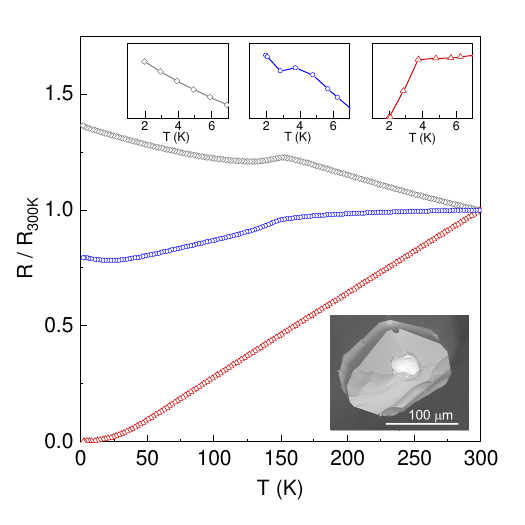}
\caption{\label{fig:Sn-core}
Three different resistivity behaviors observed in different crystals. The low temperature behaviors are shown in the upper inset. Metallic temperature dependence is correlated with a stronger superconducting character below about 4\,K (the superconducting critical temperature of Sn). The lower inset shows a scanning electron microscope image of a cross section of a broken crystal with metallic conductivity revealing a core of Sn that appears brighter in this backscatter image. In the remainder of this paper, data are only shown for crystals with resistivities that rise upon cooling and that have no anomaly associated with Sn superconductivity, like the dark gray diamonds shown here (crystal C1).
}
\end{center}
\end{figure}
\begin{figure*}
\begin{center}
\includegraphics[width=7.25in]{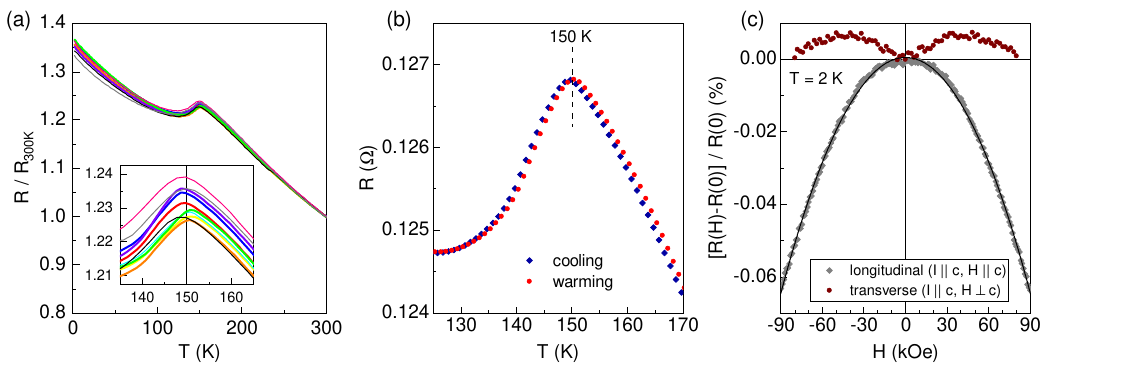}
\caption{\label{fig:res}
Results of electrical transport measurements. (a) The temperature dependence of the electrical resistance measured on ten samples with the current flowing along the \textit{c}-axis and normalized to their values at 300\,K. These include crystals C1, R28, R29, and R30 that were used to collect data shown in subsequent figures. (b) Data near the phase transition measured on heating and cooling at a rate of 2\,K per minute (crystal R28). (c) Longitudinal (crystal R29) and transverse (crystal C1) magnetoresistance measured at 2\,K, with a second order polynomial fit (black line) shown for the longitudinal data.
}
\end{center}
\end{figure*}

As noted above, some of the FeCrAs crystals grown by the method used here contained Sn-rich inclusions. Evidence suggesting this was initially seen in electrical resistance measurements, which showed widely varying temperature dependence among samples. Three examples are show in Figure \ref{fig:Sn-core}, from crystals grown by cooling in a pure Sn flux at 3.3 \degC\ per hour. The main panel shows the resistances normalized to their values at 300\,K. The behaviors range from non-metallic (an overall weak increase upon cooling), to metallic (decreasing linearly on cooling over much of the temperature range with a small residual resistance). An intermediate example is also shown. Notably, the metallic response is correlated with a sharp decrease upon cooling below about 4\,K, the superconducting transition temperature of Sn (Fig. \ref{fig:Sn-core} upper insets). A crystal with a metallic resistance curve was broken to examine its cross section in an SEM. The image is shown in the lower inset of Figure \ref{fig:Sn-core}. Indeed a Sn-rich core, appearing brighter than the bulk of the crystal in the backscatter image, was observed inside the needlelike crystal.

The presence of this Sn core seemed to be more common in the crystals with the most extreme aspect ratios, suggesting the presence of the core shell structure facilitated rapid growth along the \textit{c} direction, or vice versa. For comparison with these results, three additional growth conditions were explored, as described in the previous section. Resistance measurements were performed on several crystals from these modified growths. The statistics are limited, but it seemed that in terms of the prevalence of flux-rich cores the faster cooling rate in the pure Sn flux was similar to the slower rate, the addition of Pb to the flux exacerbated the issue, and the addition of Sb to the flux improved the odds of finding good crystals. All of the data shown in the figures that follow were collected on crystals grown from a pure Sn flux cooled at 3.3 \degC\ per hour and that had no evidence of  flux-rich cores as confirmed by R(T) measurements.

Electrical transport results are summarized in Figure \ref{fig:res}. Panel (a) shows normalized R(T) data measured on ten different crystals with the current flowing along the crystallographic \textit{c} direction, the long axis of the crystals. This includes seven crystals from a growth in a Sn flux cooled at 3.3 \degC\ per hour, one from a growth in a Sn flux cooled at 6.6 \degC\ per hour, and two from a growth in a Sn:Sb = 10:1 flux cooled at 6.6 \degC\ per hour. All show non-metallic behavior with very similar temperature dependence and no sign of contamination from the flux. These samples were used to obtain the results presented in subsequent figures.

The \textit{c}-axis resistivity of the flux-grown crystals was determined to be about $2.5 \times 10^{-4}\ \Omega$cm, similar to the value of $3.2 \times 10^{-4}\ \Omega$cm that has been reported for previous bulk crystals \cite{Wu-2009, Wu-2011}. Those studies show anisotropic behavior particularly below $T_N$; however, the needle like shape of the crystals used in the present study precluded measurement of electrical transport in directions perpendicular to the \textit{c}-axis.

The unusual temperature dependence of the electrical resistivity of this material was initially noted by Wu \textit{et al}. in Ref. \citenum{Wu-2009}, where it was discussed in the context of chemically related iron pnictide superconductor families. The resistivity generally decreases with increasing temperature from below 1\,K to over 800\,K but with a temperature dependence too weak to indicate thermal activation. Electronic structure calculations \cite{Ishida-1996,Akrap-2014} have confirmed the metallic nature of FeCrAs, with two bands crossing the Fermi level. Several ideas have been put forward to explain the non-Fermi-liquid transport behavior in this metallic material. A common theme is that the temperature dependence is dominated by scattering rather than carrier concentration. Rau and Kee have proposed that the electrons on the Fe sublattice are near a quantum critical point associated with a metal-insulator transition, giving a resistivity that increases upon cooling \cite{Rau-2011}. Akrap \textit{et al}. reported optical conductivity data that indicate an increased scattering rate as temperature is lowered \cite{Akrap-2014}. Plumb \textit{et al}. proposed that the unusual temperature dependence is related to spin fluctuations that extend up to high temperature, resulting in anomalous scattering well above the magnetic ordering temperature \cite{Plumb-2018}. In several instances \cite{Akrap-2014, Nevidomskyy-2009, Plumb-2018} the notion of a Hund's metal has been invoked, where local Hund's rule correlations lead to non-Fermi-liquid behaviors \cite{Yin-2011}. It would be interesting to see how the anomalous thermal expansion noted in the present work may inform these and other theoretical proposals for understanding the unsual electrical transport in FeCrAs.

The magnetic ordering transition in FeCrAs produces a relatively sharp cusp in the electrical resistance along the \textit{c}-axis \cite{Wu-2009, Wu-2011}. This is clearly observed in the flux-grown crystals in Figure \ref{fig:res}a. The temperature dependence between 150 and 300\,K in Figure \ref{fig:res}a is very similar to that previously reported for crystals grown from a stoichiometric melt \cite{Wu-2009, Wu-2011}. Here we see the non-metallic temperature dependence continue to 2\,K, while the previous reports show a stronger drop upon cooling below the cusp. The cusp is highlighted in Figure \ref{fig:res}b, where data collected on both heating and cooling through the transition are shown. The thermal hysteresis is limited to $<$ 1 K. Both the warming and cooling data were collected while sweeping the temperature at a rate of 2 K per minute. Although the contact between the sample and thermometer is good, some thermal lag must be experienced and would contribute to the observed small amount of hysteresis. If a component of the hysteresis is intrinsic to the sample, then it would suggest a first-order nature to the transition associated with, for example, magnetic domain formation and phase separation. The data in Figure \ref{fig:res}b indicates that if present these are weak effects. The inset of Figure \ref{fig:res}a shows that the cusp in resistance occurs at 150$\pm$1.5\,K in the flux grown crystals. In the reports on previous crystals, the cusp in the \textit{c}-axis resistivity typically occurred at 125\,K \cite{Wu-2009, Wu-2011}.

The magnetoresistance of FeCrAs was also examined. Measurements were performed with the current along the \textit{c}-axis and the magnetic field applied along the same axis (longitudinal) and also along a perpendicular direction in the \textit{ab}-plane (transverse). The results from measurements at 2\,K are shown in Figure \ref{fig:res}c. The magnitude is limited to less than 0.1\% in fields up to 90\,kOe. The transverse magnetoresistance is very small and remains positive over this field range with a nonmonotonic field dependence. The longitudinal magnetoresistance is negative and larger in magnitude with a field dependence that is close to quadratic. Ref. \citenum{Wu-2009} reports magnetoresistance from crystals grown from a stoichiometric melt and measured with the current along the \textit{a}-axis. It was found to be negative and reached a magnitude of 0.08\% at 170\,mK and 150\,kOe

\begin{figure}
\begin{center}
\includegraphics[width=3.25in]{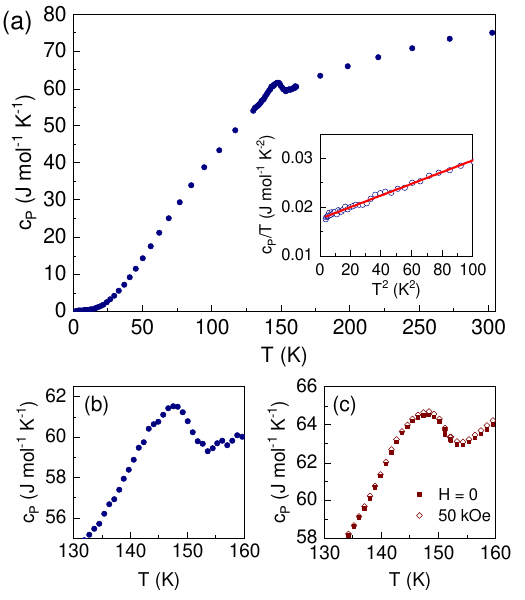}
\caption{\label{fig:hc}
Results of heat capacity measurements. (a) The temperature dependence of the heat capacity measured in zero magnetic field (crystals R28, R29, R30). The inset shows low temperature data (1.9$-$9\,K) plotted as c$_P$/T vs T$^2$ along with a linear fit. (b) Data from panel (a) near the magnetic phase transition. (c) Data from another crystal (C1) near the phase transition measured in zero field and 50\,kOe applied perpendicular to the \textit{c}-axis.
}
\end{center}
\end{figure}

Heat capacity data from flux-grown FeCrAs crystals are summarized in Figure \ref{fig:hc}. At 300\,K, values are near the expected Dulong-Petit (3R or 74.8\,J/K per mole of formula unit for FeCrAs). There is a clear anomaly near the magnetic ordering temperature, and this is highlighted in data from two different crystals in Figures \ref{fig:hc}b and \ref{fig:hc}c. A broadened lambda-like anomaly is seen. The local maximum associated with the anomaly is at 148\,K. The midpoint of the upturn upon cooling is used to define the transition temperature and is located at 150\,K. Almost no response is seen to a 50\,kOe applied magnetic field (Fig. \ref{fig:hc}c) as expected for a antiferromagnetic transition at relatively high temperature.

To estimate the amount of entropy associated with the peak in the heat capacity, a background was determined by a linear fit between the data points at 130 and 160\,K and then subtracted from the measured data. The integration of $c_P/T$ vs T using this background subtracted data gave a total entropy of 0.35 J/K/mol-FU. The small amount of entropy associated with this anomaly suggests the magnetism in FeCrAs may be itinerant with relatively small moments or that some magnetic entropy is removed by magnetic correlations that develop above $T_N$. Jin \textit{et al}. proposed itinerant magnetism with a spin density wave character for the phase transition based on the varying and relatively small Cr ordered moments of 0.8$-$1.4$\mu_B$ \cite{Jin-2019}. The magnetic excitations studied by Plumb \textit{et al}. also suggested FeCrAs is an itinerant magnet \cite{Plumb-2018}, consistent with the heat capacity behavior observed here.

The low temperature heat capacity is described well by a $T^3$ term describing the phonons and a $T$-linear term describing the free charge carriers. This is illustrated by the fit shown in the inset of Figure \ref{fig:hc}a. The fit gives a Sommerfeld coefficient of $\gamma$ = 18 mJ/K$^2$ per mole of formula unit and a Debye temperature of $\Theta_D$ = 360\,K.

Heat capacity data from FeCrAs grown from a stoichiometric melt is presented in Ref. \citenum{Wu-2009}, and some comparisons can be made to the data presented here. The most notable difference are that the authors of Ref. \citenum{Wu-2009} found only a weak anomaly associated with magnetic ordering at 125\,K, while here we see a clear peak at 150\,K. In addition, Ref. \citenum{Wu-2009} reports a higher Sommerfeld coefficient (30 mJ/K$^2$ per mole of formula unit), while the Debye temperature appears to be roughly consistent with the value determined here.

\begin{figure*}
\begin{center}
\includegraphics[width=7.25in]{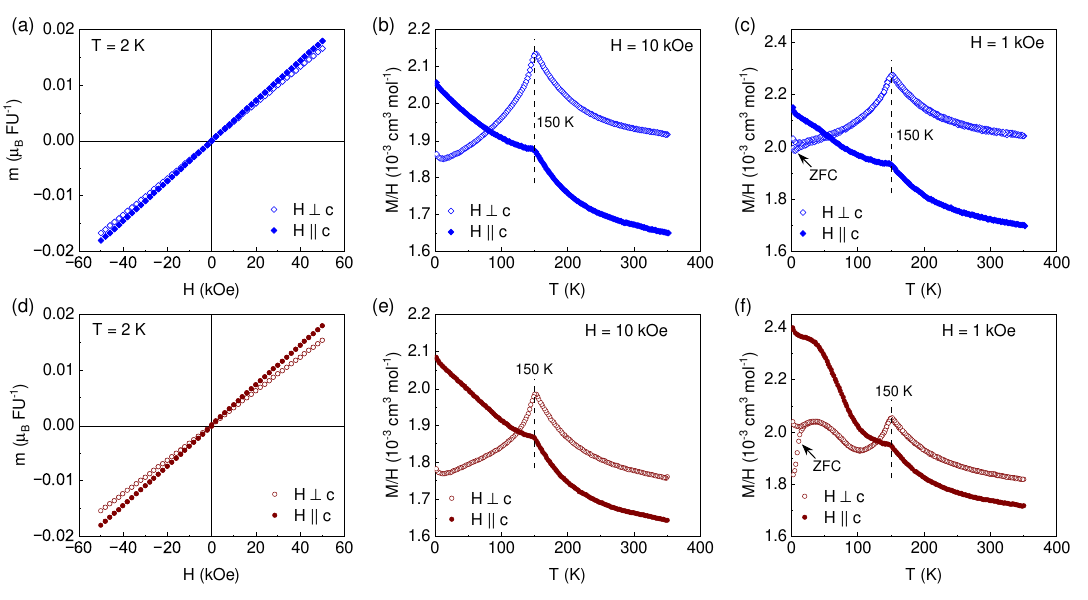}
\caption{\label{fig:mag}
Results of magnetization measurements for two different samples (crystal C1 in panels a, b, and c, and crystals R28, R29, and R30 measured together in panels d, e, and f). Isothermal magnetization curves measured at 2 K are shown in panels (a) and (d). Temperature dependent data measured at 10 kOe are shown in panels (b) and (e). Temperature dependent data measured at 1 kOe are shown in panels (c) and (f), with both field-cooled-cooling and zero-field-cooled-warming (ZFC) data shown for $H \perp c$.
}
\end{center}
\end{figure*}

As noted above, early studies of the magnetism in FeCrAs included a powder neutron diffraction study \cite{Swainson-2010}. Analysis of the data collected at 2.8\,K indicated a $3\times 3\times 1$ magnetic supercell [$k = (\frac{1}{3}, \frac{1}{3}, 0$)], with magnetic moments lying in the \textit{ab}-plane on Cr sites. The data indicated no ordered magnetic moment on Fe. This interesting and counterintuitive result had been predicted from electronic structure calculations by Ishida \textit{et al}. \cite{Ishida-1996}. Rau and Kee have proposed a model Hamiltonian for this material that supports a spin-liquid phase on the Fe sublattice, consistent with the lack of an ordered Fe moment \cite{Rau-2011}. This model also provides a source of fluctuations consistent with the observed transport properties and heat capacity of FeCrAs. The magnetic structure was later revisited using single crystal neutron diffraction \cite{Jin-2019}. The propagation vector of $k = (\frac{1}{3}, \frac{1}{3}, 0$) with noncollinear Cr spins lying in the \textit{ab}-plane was confirmed in the ground state (at 2.5\,K). A spiral-like structure for the Cr moments was determined with moment sizes varying from 0.8 to 1.4\,$\mu_B$, and the ordered Fe moments were constrained to be less than 0.1\,$\mu_B$. A spin reorientation from in-plane at low temperature to along \textit{c} near $T_N$ was indicated, but it is not clear that this is common to all samples \cite{Jin-2019}. The bulk magnetic order was also probed using muon spin relaxation and magnetic x-ray scattering, with results consistent with the non-collinear magnetic structure involving ordered moments only at the Cr sites \cite{Huddart-2019}. It is also noteworthy that, the noncollinear magnetic structure and the symmetry of the crystal lattice have raised interest in FeCrAs as a candidate altermagnet \cite{Singh-2025}.

The temperature and field dependence of the magnetic susceptibility of flux-grown FeCrAs is summarized in Figure \ref{fig:mag}. Results from two samples are presented. Panels a-c show data from a single crystal, labeled C1. This is the crystal used to collect the heat capacity data shown in Figure \ref{fig:hc}c. Panels d-f show data from a collection of three crystals, labeled R28, R29, and R30. These three crystals made up the sample used for to collect the heat capacity data shown in Figure \ref{fig:hc}a and \ref{fig:hc}b. Resistivity data for all four of these crystals are included in Figure \ref{fig:res}a.

Isothermal magnetization curves were linear up to 50\,kOe at all temperatures investigated. Data collected at 2\,K are shown in Figure \ref{fig:mag}a and \ref{fig:mag}d. The magnetic phase transition is apparent in the temperature dependent data as a cusp that occurs at 150\,K (Fig. \ref{fig:mag}(b, c, e, f)). When the field is applied perpendicular to the \textit{c}-axis, the magnetization upon cooling drops sharply below $T_N$. When the field is applied along the \textit{c}-axis, the magnetization continues to rise upon cooling below the cusp. In the simplest interpretation, this suggests that the ordered moments lie in the \textit{ab}-plane. This is consistent with neutron diffraction results discussed above \cite{Swainson-2010, Jin-2019}. Note the that overall temperature dependence is weak, with the susceptibility changing by less than a factor of 1.4 between 2 and 350\,K, and by only about 10\% between $T_N$ and $2T_N$. This, like the heat capacity, is consistent with itinerant magnetism in FeCrAs.

In addition to the cusp at $T_N$, other notable features in the temperature dependence of the susceptibility include a small upturn upon cooling at the lowest temperatures, and divergence between field-cooled and zero-field-cooled data. The divergence is stronger in Figure \ref{fig:mag}f than in Figure \ref{fig:mag}c, suggesting it may be related to the broad hump-like feature seen below $T_N$ in the latter. This likely indicates some glassiness perhaps related to defects, or a freezing of domains that do not have fully compensated moments at low temperature.

The behavior of the magnetic susceptibility described above is generally consistent with previous reports. Similar anisotropic susceptibility behavior was reported by Wu \textit{et al}. \cite{Wu-2009, Wu-2011} Jin \textit{et al}. \cite{Jin-2019}. In those reports and in Ref. \citenum{Huddart-2019} some glassiness at low temperature was also noted. The temperature and strength of the divergence between field-cooled and zero-field-cooled data was observed to vary from sample to sample in the literature, as it does between the samples shown in Figure \ref{fig:mag}c and \ref{fig:mag}f here. This suggests that this glassy behavior may be related to defects. The sample studied in Ref. \citenum{Jin-2019} showed an additional phase transition below $T_N$. This was confirmed with both magnetization measurements and neutron diffraction and is attributed to canting of the ordered Cr moments, but this transition is not seen in other reports.

As noted above in the discussion of heat capacity and electrical transport data, a key difference between the flux-grown crystals and previous samples is value of $T_N$. Here we find a value of 150\,K compared to 100$-$125\,K \cite{Wu-2009, Wu-2011, Plumb-2018, Huddart-2019, Jin-2019}. The other key differences noted above are the value of the Sommerfeld coefficient and shape/size of the heat capacity anomaly. These differences may reflect a difference in stoichiometry or degree of Cr/Fe site ordering between the crystals used in the various studies. Based on the early report of a partial solid solution range in the \ce{Fe2As} and \ce{Cr2As} pseudobinary phase diagram \cite{Hollan-1966}, some proclivity toward Fe/Cr mixing on the two transition metal sites should be assumed, and entropy would favor this. At a simple level, if only the Cr electrons participate in the magnetic ordering, then increasing the Cr content may be expected to strengthen the magnetic order and reduce the free carrier concentration at low temperature, consistent with increased $T_N$ and reduced $\gamma$ seen here.

\section{Summary and Conclusions}

The FeCrAs crystals studied here, grown from a Sn flux, show an antiferromagnetic ordering transition like that previously reported for crystals grown from a stoichiometric melt \cite{Wu-2009, Swainson-2010, Wu-2011}. However, there are notable differences in the physical properties of the crystals. First, the N\'{e}el temperature is found to be 150\,K and to show very little variation from crystal to crystal, including crystals grown under different conditions (cooling rate and flux composition). This is about 25\,K higher than in previous reports, which also showed more variation in $T_N$ \cite{Wu-2009, Wu-2011, Plumb-2018, Huddart-2019, Jin-2019}. In addition, the Sommerfeld coefficient of flux grown FeCrAs is found to be 18\,mJ/K$^2$/mol, a little more than half of the previously reported value \cite{Wu-2009}. Semiquantitative analysis suggests the crystals are Cr-rich, with a composition near Fe$_{0.9}$Cr$_{1.1}$As, which may be correlated to the differences in observed magnetic ordering temperatures and Sommerfeld coefficients. It has been reported that the magnetism arises from Cr electrons, while the Fe electrons are expected to remain itinerant \cite{Swainson-2010, Huddart-2019}. Thus, it may be expected that increasing the Cr content relative to Fe should strengthen the magnetism and reduce the density of states contributing to conduction.  The present work also highlights the care that must be taken when interpreting transport data from crystals grown in this way, due to the possibility of flux inclusions.

Temperature dependent structural data reveal unusual thermal expansion behavior suggesting significant coupling between the magnetism and the lattice. Nearly temperature independent thermal expansion along the hexagonal \textit{c}-axis is observed above the N\'{e}el temperature. This behavior deserves further study, and may provide insight into non-metallic transport properties associated with anomalous scattering observed over a wide temperature range in FeCrAs. The variation in properties between crystals grown by different techniques relates back to the phase width that this compound is expected to have \cite{Hollan-1966}. Careful control of the stoichiometry and associated Fe/Cr mixing in the crystals may be essential in understanding the magnetism, transport, and spin-lattice coupling in this interesting material, and flux growths like those described here provide a good approach to these types of studies.

\section{Methods}

The crystal growth is described in the main text. Single crystal x-ray diffraction data were collected using a Bruker D8 Advance Quest diffractometer with a graphite monochromator and Mo-K$\alpha$ radiation. For these measurements, shards broken from the needlelike crystals were mounted on the edge of a kapton loop with a small amount of Parabar oil. Data were reduced using tools in the Bruker Apex-4 software and structures were refined using ShelX. The chemical compositions were determined using semiquantitative analysis of energy dispersive spectroscopy data collected with a Oxford Aztec spectrometer on a Hitachi TM-4000Plus tabletop scanning electron microscope.

Electrical transport and heat capacity measurements were performed using Physical Property Measurement Systems (PPMS) from Quantum Design employing standard options and practices. Electrical contacts were made using silver epoxy. Magnetization data was collected using a Magnetic Property Measurement System (MPMS-3) from Quantum Design, with the crystals mounted on quartz paddle sample holders using GE varnish.

Multiple crystals were used to collect the physical property data in this work. All measurements were done on samples grown using the first growth procedure described in the text, cooling at 3.3 \degC\ per hour with a pure Sn flux, with the exception of some of the resistivity measurements shown in Figure \ref{fig:res}a only. Some crystals were used for multiple measurements. Crystal labels are indicated in the figure captions, and summarized here. Crystals C1, R28, R29, and R30 are the sources of data for all panels except Figure \ref{fig:Sn-core}, which includes data from two unnamed Sn-contaminated crystals, and Figure \ref{fig:res}a, which includes data from these four as well as multiple other clean crystals.  In Figure \ref{fig:res}a, data from C1, R28, R29, and R30 appear as yellow, red, blue, and cyan lines, respectively. Data from crystal C1 also appears in Figure \ref{fig:Sn-core} (gray diamonds), Figure \ref{fig:res}c (maroon circles), Figure \ref{fig:hc}c, and Figure \ref{fig:mag}a-c. Data from crystal R28 appears in Figure \ref{fig:res}b. Data from crystal R29 appears in Figure \ref{fig:res}c (gray diamonds). Data from crystals R28, R29, and R30 appear as collectively as a single sample in Figures \ref{fig:hc}a, \ref{fig:hc}b, and \ref{fig:mag}d-f.

\section{Acknowledgements}
We thank Norman Mannella and Paolo Vilmercati for directing our attention to this interesting material. This work was supported by the U.S. Department of Energy, Office of Science, Basic Energy Sciences, Materials Sciences and Engineering Division.

\section{Supporting Information}
Crystallographic information files for the structural refinements at 100 and 200\,K.

\bibliography{FeCrAs-bib}

\providecommand{\latin}[1]{#1}
\makeatletter
\providecommand{\doi}
  {\begingroup\let\do\@makeother\dospecials
  \catcode`\{=1 \catcode`\}=2 \doi@aux}
\providecommand{\doi@aux}[1]{\endgroup\texttt{#1}}
\makeatother
\providecommand*\mcitethebibliography{\thebibliography}
\csname @ifundefined\endcsname{endmcitethebibliography}
  {\let\endmcitethebibliography\endthebibliography}{}
\begin{mcitethebibliography}{39}
\providecommand*\natexlab[1]{#1}
\providecommand*\mciteSetBstSublistMode[1]{}
\providecommand*\mciteSetBstMaxWidthForm[2]{}
\providecommand*\mciteBstWouldAddEndPuncttrue
  {\def\EndOfBibitem{\unskip.}}
\providecommand*\mciteBstWouldAddEndPunctfalse
  {\let\EndOfBibitem\relax}
\providecommand*\mciteSetBstMidEndSepPunct[3]{}
\providecommand*\mciteSetBstSublistLabelBeginEnd[3]{}
\providecommand*\EndOfBibitem{}
\mciteSetBstSublistMode{f}
\mciteSetBstMaxWidthForm{subitem}{(\alph{mcitesubitemcount})}
\mciteSetBstSublistLabelBeginEnd
  {\mcitemaxwidthsubitemform\space}
  {\relax}
  {\relax}

\bibitem[Villars and Cenzual(2025)Villars, and Cenzual]{Villars-2024}
Villars,~P.; Cenzual,~K. {Pearson's Crystal Data - Crystal Structure Database
  for Inorganic Compounds}. Release 2024/25, 2025\relax
\mciteBstWouldAddEndPuncttrue
\mciteSetBstMidEndSepPunct{\mcitedefaultmidpunct}
{\mcitedefaultendpunct}{\mcitedefaultseppunct}\relax
\EndOfBibitem
\bibitem[Dshemuchadse and Steurer(2015)Dshemuchadse, and
  Steurer]{Dshemuchadse-2015}
Dshemuchadse,~J.; Steurer,~W. Some Statistics on Intermetallic Compounds.
  \emph{Inorg. Chem.} \textbf{2015}, \emph{54}, 1120\relax
\mciteBstWouldAddEndPuncttrue
\mciteSetBstMidEndSepPunct{\mcitedefaultmidpunct}
{\mcitedefaultendpunct}{\mcitedefaultseppunct}\relax
\EndOfBibitem
\bibitem[Fruchart(1982)]{Fruchart-1982}
Fruchart,~R. Effets d’{\'e}lectron{\'e}gativit{\'e} et interactions
  m{\'e}talliques dans les phosphures et ars{\'e}niures ternaires des
  {\'e}l{\'e}ments de transition 3d, 4d, 5d de type m{\'e}tallique. \emph{Ann.
  Chim.} \textbf{1982}, \emph{7}, 563\relax
\mciteBstWouldAddEndPuncttrue
\mciteSetBstMidEndSepPunct{\mcitedefaultmidpunct}
{\mcitedefaultendpunct}{\mcitedefaultseppunct}\relax
\EndOfBibitem
\bibitem[Ku{\ss}mann \latin{et~al.}(1998)Ku{\ss}mann, P\"{o}ttgen, K\"{u}nnen,
  Kotzyba, M\"{u}llmann, and Mosel]{Kussmann-1998}
Ku{\ss}mann,~D.; P\"{o}ttgen,~R.; K\"{u}nnen,~B.; Kotzyba,~G.;
  M\"{u}llmann,~R.; Mosel,~B.~D. Dimorphic YbPdSn with ZrNiAl and TiNiSi type
  structure. \emph{Z. Kristallogr.} \textbf{1998}, \emph{213}, 356\relax
\mciteBstWouldAddEndPuncttrue
\mciteSetBstMidEndSepPunct{\mcitedefaultmidpunct}
{\mcitedefaultendpunct}{\mcitedefaultseppunct}\relax
\EndOfBibitem
\bibitem[Oliynyk \latin{et~al.}(2017)Oliynyk, Adutwum, Rudyk, Pisavadia, Lotfi,
  Hlukhyy, Harynuk, Mar, and Brgoch]{Oliynyk-2025}
Oliynyk,~A.~O.; Adutwum,~L.~A.; Rudyk,~B.~W.; Pisavadia,~H.; Lotfi,~S.;
  Hlukhyy,~V.; Harynuk,~J.~J.; Mar,~A.; Brgoch,~J. Disentangling Structural
  Confusion through Machine Learning: Structure Prediction and Polymorphism of
  Equiatomic Ternary Phases ABC. \emph{J. Am. Chem. Soc.} \textbf{2017},
  \emph{139}, 17870--17881\relax
\mciteBstWouldAddEndPuncttrue
\mciteSetBstMidEndSepPunct{\mcitedefaultmidpunct}
{\mcitedefaultendpunct}{\mcitedefaultseppunct}\relax
\EndOfBibitem
\bibitem[Bao \latin{et~al.}(2021)Bao, Bugaris, Zheng, Chung, and
  Kanatzidis]{Bao-2021}
Bao,~J.-K.; Bugaris,~D.~E.; Zheng,~H.; Chung,~D.~Y.; Kanatzidis,~M.~G. A
  Noncentrosymmetric Polymorph of LuRuGe. \emph{Inorg. Chem.} \textbf{2021},
  \emph{60}, 7827\relax
\mciteBstWouldAddEndPuncttrue
\mciteSetBstMidEndSepPunct{\mcitedefaultmidpunct}
{\mcitedefaultendpunct}{\mcitedefaultseppunct}\relax
\EndOfBibitem
\bibitem[Kong \latin{et~al.}(2024)Kong, Singh, Sarkar, Viswanathan, Kolen’ko,
  Mudryk, Johnson, and Kovnir]{Kong-2024}
Kong,~S.; Singh,~P.; Sarkar,~A.; Viswanathan,~G.; Kolen’ko,~Y.~V.;
  Mudryk,~Y.; Johnson,~D.~D.; Kovnir,~K. Enhancing Properties with Distortion:
  A Comparative Study of Two Iron Phosphide Fe2P Polymorphs. \emph{Chem.
  Mater.} \textbf{2024}, \emph{36}, 1665\relax
\mciteBstWouldAddEndPuncttrue
\mciteSetBstMidEndSepPunct{\mcitedefaultmidpunct}
{\mcitedefaultendpunct}{\mcitedefaultseppunct}\relax
\EndOfBibitem
\bibitem[Roy-Montreuil \latin{et~al.}(1972)Roy-Montreuil, Deyris, Michael,
  Rouault, l'H\'{e}ritier, Nylund, S\'{e}nateur, and
  Fruchart]{Roy-Montreuil-1972}
Roy-Montreuil,~M.; Deyris,~B.; Michael,~A.; Rouault,~A.; l'H\'{e}ritier,~P.;
  Nylund,~A.; S\'{e}nateur,~J.~P.; Fruchart,~R. Nouveaux Composes Ternaires
  MM'P et MM'As Interactions Metalliques et Structures. \emph{Mater. Res.
  Bull.} \textbf{1972}, \emph{7}, 813\relax
\mciteBstWouldAddEndPuncttrue
\mciteSetBstMidEndSepPunct{\mcitedefaultmidpunct}
{\mcitedefaultendpunct}{\mcitedefaultseppunct}\relax
\EndOfBibitem
\bibitem[Meier \latin{et~al.}(2020)Meier, Du, Okamoto, Mohanta, May, McGuire,
  Bridges, Samolyuk, and Sales]{Meier-2020}
Meier,~W.~R.; Du,~M.-H.; Okamoto,~S.; Mohanta,~N.; May,~A.~F.; McGuire,~M.~A.;
  Bridges,~C.~A.; Samolyuk,~G.~D.; Sales,~B.~C. {Flat bands in the CoSn-type
  compounds}. \emph{Phys. Rev. B} \textbf{2020}, \emph{102}, 075148\relax
\mciteBstWouldAddEndPuncttrue
\mciteSetBstMidEndSepPunct{\mcitedefaultmidpunct}
{\mcitedefaultendpunct}{\mcitedefaultseppunct}\relax
\EndOfBibitem
\bibitem[Ptok \latin{et~al.}(2021)Ptok, Kobia\l{}ka, Sternik,
  \L{}a\ifmmode~\dot{z}\else \.{z}\fi{}ewski, Jochym,
  Ole\ifmmode~\acute{s}\else \'{s}\fi{}, Stankov, and Piekarz]{Ptok-2021}
Ptok,~A.; Kobia\l{}ka,~A.; Sternik,~M.; \L{}a\ifmmode~\dot{z}\else
  \.{z}\fi{}ewski,~J.; Jochym,~P.~T.; Ole\ifmmode~\acute{s}\else
  \'{s}\fi{},~A.~M.; Stankov,~S.; Piekarz,~P. {Chiral phonons in the honeycomb
  sublattice of layered CoSn-like compounds}. \emph{Phys. Rev. B}
  \textbf{2021}, \emph{104}, 054305\relax
\mciteBstWouldAddEndPuncttrue
\mciteSetBstMidEndSepPunct{\mcitedefaultmidpunct}
{\mcitedefaultendpunct}{\mcitedefaultseppunct}\relax
\EndOfBibitem
\bibitem[Ptok \latin{et~al.}(2023)Ptok, Meier, Kobia\l{}ka, Basak, Sternik,
  \L{}a\ifmmode~\dot{z}\else \.{z}\fi{}ewski, Jochym, McGuire, Sales, Miao,
  Piekarz, and Ole\ifmmode~\acute{s}\else \'{s}\fi{}]{Ptok-2023}
Ptok,~A.; Meier,~W.~R.; Kobia\l{}ka,~A.; Basak,~S.; Sternik,~M.;
  \L{}a\ifmmode~\dot{z}\else \.{z}\fi{}ewski,~J.; Jochym,~P.~T.;
  McGuire,~M.~A.; Sales,~B.~C.; Miao,~H.; Piekarz,~P.;
  Ole\ifmmode~\acute{s}\else \'{s}\fi{},~A.~M. {Phononic drumhead surface state
  in the distorted kagome compound RhPb}. \emph{Phys. Rev. Res.} \textbf{2023},
  \emph{5}, 043231\relax
\mciteBstWouldAddEndPuncttrue
\mciteSetBstMidEndSepPunct{\mcitedefaultmidpunct}
{\mcitedefaultendpunct}{\mcitedefaultseppunct}\relax
\EndOfBibitem
\bibitem[Barz \latin{et~al.}(1980)Barz, Ku, Meisner, Fisk, and
  Matthias]{Barz-1980}
Barz,~H.; Ku,~H.; Meisner,~G.; Fisk,~Z.; Matthias,~B. {Ternary transition metal
  phosphides: High-temperature superconductors}. \emph{Proc. Natl. Acad. Sci.
  USA.} \textbf{1980}, \emph{77}, 3132--3134\relax
\mciteBstWouldAddEndPuncttrue
\mciteSetBstMidEndSepPunct{\mcitedefaultmidpunct}
{\mcitedefaultendpunct}{\mcitedefaultseppunct}\relax
\EndOfBibitem
\bibitem[Inohara \latin{et~al.}(2016)Inohara, Okamoto, Yamakawa, and
  Takenaka]{Inohara-2016}
Inohara,~T.; Okamoto,~Y.; Yamakawa,~Y.; Takenaka,~K. {Synthesis and
  superconducting properties of a hexagonal phosphide ScRhP}. \emph{J. Phys.
  Soc. Jpn.} \textbf{2016}, \emph{85}, 094706\relax
\mciteBstWouldAddEndPuncttrue
\mciteSetBstMidEndSepPunct{\mcitedefaultmidpunct}
{\mcitedefaultendpunct}{\mcitedefaultseppunct}\relax
\EndOfBibitem
\bibitem[Kumar \latin{et~al.}(2022)Kumar, Luo, Du, Su, Zhang, Cao, and
  Yuan]{Kumar-2022}
Kumar,~R.; Luo,~S.-S.; Du,~F.; Su,~H.; Zhang,~J.; Cao,~C.; Yuan,~H.
  {Superconductivity in non-centrosymmetric ZrNiAl and HfRhSn-type compounds}.
  \emph{J. Phys. Condens. Matter.} \textbf{2022}, \emph{34}, 435701\relax
\mciteBstWouldAddEndPuncttrue
\mciteSetBstMidEndSepPunct{\mcitedefaultmidpunct}
{\mcitedefaultendpunct}{\mcitedefaultseppunct}\relax
\EndOfBibitem
\bibitem[Su \latin{et~al.}(2021)Su, Shang, Du, Chen, Ye, Lu, Cao, Smidman, and
  Yuan]{Su-2021}
Su,~H.; Shang,~T.; Du,~F.; Chen,~C.~F.; Ye,~H.~Q.; Lu,~X.; Cao,~C.;
  Smidman,~M.; Yuan,~H.~Q. {NbReSi: A noncentrosymetric superconductor with
  large upper critical field}. \emph{Phys. Rev. Mater.} \textbf{2021},
  \emph{5}, 114802\relax
\mciteBstWouldAddEndPuncttrue
\mciteSetBstMidEndSepPunct{\mcitedefaultmidpunct}
{\mcitedefaultendpunct}{\mcitedefaultseppunct}\relax
\EndOfBibitem
\bibitem[Xie \latin{et~al.}(2022)Xie, Du, Zheng, Su, Nie, Liu, Xia, Shang, Cao,
  Smidman, Takabatake, and Yuan]{Xie-2022}
Xie,~W.; Du,~F.; Zheng,~X.~Y.; Su,~H.; Nie,~Z.~Y.; Liu,~B.~Q.; Xia,~Y.~H.;
  Shang,~T.; Cao,~C.; Smidman,~M.; Takabatake,~T.; Yuan,~H.~Q. {Semimetallic
  Kondo lattice behavior in YbPdAs with a distorted kagome structure}.
  \emph{Phys. Rev. B} \textbf{2022}, \emph{106}, 075132\relax
\mciteBstWouldAddEndPuncttrue
\mciteSetBstMidEndSepPunct{\mcitedefaultmidpunct}
{\mcitedefaultendpunct}{\mcitedefaultseppunct}\relax
\EndOfBibitem
\bibitem[Laha \latin{et~al.}(2020)Laha, Mardanya, Singh, Lin, Bansil, Agarwal,
  and Hossain]{Laha-2020}
Laha,~A.; Mardanya,~S.; Singh,~B.; Lin,~H.; Bansil,~A.; Agarwal,~A.;
  Hossain,~Z. {Magnetotransport properties of the topological nodal-line
  semimetal CaCdSn}. \emph{Phys. Rev. B} \textbf{2020}, \emph{102},
  035164\relax
\mciteBstWouldAddEndPuncttrue
\mciteSetBstMidEndSepPunct{\mcitedefaultmidpunct}
{\mcitedefaultendpunct}{\mcitedefaultseppunct}\relax
\EndOfBibitem
\bibitem[Wu \latin{et~al.}(2009)Wu, McCollam, Swainson, Rancourt, and
  Julian]{Wu-2009}
Wu,~W.; McCollam,~A.; Swainson,~I.; Rancourt,~D.; Julian,~S. {A novel
  non-Fermi-liquid state in the iron-pnictide FeCrAs}. \emph{Europhys. Lett.}
  \textbf{2009}, \emph{85}, 17009\relax
\mciteBstWouldAddEndPuncttrue
\mciteSetBstMidEndSepPunct{\mcitedefaultmidpunct}
{\mcitedefaultendpunct}{\mcitedefaultseppunct}\relax
\EndOfBibitem
\bibitem[Swainson \latin{et~al.}(2010)Swainson, Wu, McCollam, and
  Julian]{Swainson-2010}
Swainson,~I.~P.; Wu,~W.; McCollam,~A.; Julian,~S.~R. {Non-collinear
  antiferromagnetism in FeCrAs}. \emph{Can. J. Phys.} \textbf{2010}, \emph{88},
  701--706\relax
\mciteBstWouldAddEndPuncttrue
\mciteSetBstMidEndSepPunct{\mcitedefaultmidpunct}
{\mcitedefaultendpunct}{\mcitedefaultseppunct}\relax
\EndOfBibitem
\bibitem[Wu \latin{et~al.}(2011)Wu, McCollam, Swainson, and Julian]{Wu-2011}
Wu,~W.~L.; McCollam,~A.; Swainson,~I.~P.; Julian,~S.~R. {Crystal growth,
  structure, and incoherent metallic behaviour of FeCrAs}. \emph{Solid State
  Phenomena} \textbf{2011}, \emph{170}, 276--281\relax
\mciteBstWouldAddEndPuncttrue
\mciteSetBstMidEndSepPunct{\mcitedefaultmidpunct}
{\mcitedefaultendpunct}{\mcitedefaultseppunct}\relax
\EndOfBibitem
\bibitem[Jin \latin{et~al.}(2019)Jin, Meven, Deng, Su, Wu, Julian, and
  Kim]{Jin-2019}
Jin,~W.; Meven,~M.; Deng,~H.; Su,~Y.; Wu,~W.; Julian,~S.; Kim,~Y.-J. {Spin
  reorientation in FeCrAs revealed by single-crystal neutron diffraction}.
  \emph{Phys. Rev. B} \textbf{2019}, \emph{100}, 174421\relax
\mciteBstWouldAddEndPuncttrue
\mciteSetBstMidEndSepPunct{\mcitedefaultmidpunct}
{\mcitedefaultendpunct}{\mcitedefaultseppunct}\relax
\EndOfBibitem
\bibitem[Huddart \latin{et~al.}(2019)Huddart, Birch, Pratt, Blundell, Porter,
  Clark, Wu, Julian, Hatton, and Lancaster]{Huddart-2019}
Huddart,~B.; Birch,~M.~T.; Pratt,~F.; Blundell,~S.; Porter,~D.~G.;
  Clark,~S.~J.; Wu,~W.; Julian,~S.~R.; Hatton,~P.; Lancaster,~T. {Local
  magnetism, magnetic order and spin freezing in the ‘nonmetallic metal’
  FeCrAs}. \emph{J. Phys. Condens. Matter} \textbf{2019}, \emph{31},
  285803\relax
\mciteBstWouldAddEndPuncttrue
\mciteSetBstMidEndSepPunct{\mcitedefaultmidpunct}
{\mcitedefaultendpunct}{\mcitedefaultseppunct}\relax
\EndOfBibitem
\bibitem[Rancourt(1984)]{rancourt1984hyperfine}
Rancourt,~D.~G. Hyperfine field fluctuations in the M\"{o}ssbauer spectrum of
  magnetic materials: Application to small particles and to the bulk
  antiferromagnetic. PhD Thesis, University of Toronto, 1984\relax
\mciteBstWouldAddEndPuncttrue
\mciteSetBstMidEndSepPunct{\mcitedefaultmidpunct}
{\mcitedefaultendpunct}{\mcitedefaultseppunct}\relax
\EndOfBibitem
\bibitem[Ishida \latin{et~al.}(1996)Ishida, Takiguchi, Fujii, and
  Asano]{Ishida-1996}
Ishida,~S.; Takiguchi,~T.; Fujii,~S.; Asano,~S. {Magnetic properties and
  electronic structures of CrMZ (M = Fe, Co, Ni; Z = P, As)}. \emph{Phys. B
  Condens. Matt.} \textbf{1996}, \emph{217}, 87\relax
\mciteBstWouldAddEndPuncttrue
\mciteSetBstMidEndSepPunct{\mcitedefaultmidpunct}
{\mcitedefaultendpunct}{\mcitedefaultseppunct}\relax
\EndOfBibitem
\bibitem[Rau and Kee(2011)Rau, and Kee]{Rau-2011}
Rau,~J.~G.; Kee,~H.-Y. {Hidden spin liquid in an antiferromagnet: Applications
  to FeCrAs}. \emph{Phys. Rev. B} \textbf{2011}, \emph{84}, 104448\relax
\mciteBstWouldAddEndPuncttrue
\mciteSetBstMidEndSepPunct{\mcitedefaultmidpunct}
{\mcitedefaultendpunct}{\mcitedefaultseppunct}\relax
\EndOfBibitem
\bibitem[Akrap \latin{et~al.}(2014)Akrap, Dai, Wu, Julian, and
  Homes]{Akrap-2014}
Akrap,~A.; Dai,~Y.; Wu,~W.; Julian,~S.; Homes,~C. {Optical properties and
  electronic structure of the nonmetallic metal FeCrAs}. \emph{Phys. Rev. B}
  \textbf{2014}, \emph{89}, 125115\relax
\mciteBstWouldAddEndPuncttrue
\mciteSetBstMidEndSepPunct{\mcitedefaultmidpunct}
{\mcitedefaultendpunct}{\mcitedefaultseppunct}\relax
\EndOfBibitem
\bibitem[Plumb \latin{et~al.}(2018)Plumb, Stock, Rodriguez-Rivera, Castellan,
  Taylor, Lau, Wu, Julian, and Kim]{Plumb-2018}
Plumb,~K.; Stock,~C.; Rodriguez-Rivera,~J.; Castellan,~J.-P.; Taylor,~J.;
  Lau,~B.; Wu,~W.; Julian,~S.; Kim,~Y.-J. {From mean-field localized magnetism
  to itinerant spin fluctuations in the ``nonmetallic metal'' FeCrAs}.
  \emph{Phys. Rev. B} \textbf{2018}, \emph{97}, 184431\relax
\mciteBstWouldAddEndPuncttrue
\mciteSetBstMidEndSepPunct{\mcitedefaultmidpunct}
{\mcitedefaultendpunct}{\mcitedefaultseppunct}\relax
\EndOfBibitem
\bibitem[Singh \latin{et~al.}(2025)Singh, Cheong, and Guo]{Singh-2025}
Singh,~D.~K.; Cheong,~S.-W.; Guo,~J. {Altermagnetism in NiSi and
  Antiferromagnetic Candidate Materials with Non-Collinear Spins}. \emph{Adv.
  Phys. Res.} \textbf{2025}, \emph{n/a}, 2400170\relax
\mciteBstWouldAddEndPuncttrue
\mciteSetBstMidEndSepPunct{\mcitedefaultmidpunct}
{\mcitedefaultendpunct}{\mcitedefaultseppunct}\relax
\EndOfBibitem
\bibitem[Kanatzidis \latin{et~al.}(2005)Kanatzidis, P{\"o}ttgen, and
  Jeitschko]{Kanatzidis-2005}
Kanatzidis,~M.~G.; P{\"o}ttgen,~R.; Jeitschko,~W. The metal flux: a preparative
  tool for the exploration of intermetallic compounds. \emph{Angew. Chem.}
  \textbf{2005}, \emph{44}, 6996--7023\relax
\mciteBstWouldAddEndPuncttrue
\mciteSetBstMidEndSepPunct{\mcitedefaultmidpunct}
{\mcitedefaultendpunct}{\mcitedefaultseppunct}\relax
\EndOfBibitem
\bibitem[Katsuraki and Achiwa(1966)Katsuraki, and Achiwa]{Katsuraki-1966}
Katsuraki,~H.; Achiwa,~N. {The magnetic structure of Fe$_2$As}. \emph{J. Phys.
  Soc. Jpn.} \textbf{1966}, \emph{21}, 2238--2243\relax
\mciteBstWouldAddEndPuncttrue
\mciteSetBstMidEndSepPunct{\mcitedefaultmidpunct}
{\mcitedefaultendpunct}{\mcitedefaultseppunct}\relax
\EndOfBibitem
\bibitem[Canfield \latin{et~al.}(2016)Canfield, Kong, Kaluarachchi, and
  Jo]{Canfield-2016}
Canfield,~P.~C.; Kong,~T.; Kaluarachchi,~U.~S.; Jo,~N.~H. Use of frit-disc
  crucibles for routine and exploratory solution growth of single crystalline
  samples. \emph{Philos. Mag.} \textbf{2016}, \emph{96}, 84--92\relax
\mciteBstWouldAddEndPuncttrue
\mciteSetBstMidEndSepPunct{\mcitedefaultmidpunct}
{\mcitedefaultendpunct}{\mcitedefaultseppunct}\relax
\EndOfBibitem
\bibitem[Hollan(1966)]{Hollan-1966}
Hollan,~L. {\'{E}tude structurale et magn\'{e}tique d'ars\'{e}niures mixtes
  {M$_2$As}}. \emph{Ann. Chim. (Paris)} \textbf{1966}, \emph{1}, 437\relax
\mciteBstWouldAddEndPuncttrue
\mciteSetBstMidEndSepPunct{\mcitedefaultmidpunct}
{\mcitedefaultendpunct}{\mcitedefaultseppunct}\relax
\EndOfBibitem
\bibitem[Song \latin{et~al.}(2021)Song, Shi, Deng, Xing, and Chen]{Song2021}
Song,~Y.; Shi,~N.; Deng,~S.; Xing,~X.; Chen,~J. Negative thermal expansion in
  magnetic materials. \emph{Prog. Mater. Sci.} \textbf{2021}, \emph{121},
  100835\relax
\mciteBstWouldAddEndPuncttrue
\mciteSetBstMidEndSepPunct{\mcitedefaultmidpunct}
{\mcitedefaultendpunct}{\mcitedefaultseppunct}\relax
\EndOfBibitem
\bibitem[Zach \latin{et~al.}(1990)Zach, Guillot, and Fruchart]{Zach1990}
Zach,~R.; Guillot,~M.; Fruchart,~R. {The influence of high magnetic fields on
  the first order magneto-elastic transition in MnFe(P$_{1- y}$As$_y$)
  systems}. \emph{J. Magn. Magn. Mater.} \textbf{1990}, \emph{89},
  221--228\relax
\mciteBstWouldAddEndPuncttrue
\mciteSetBstMidEndSepPunct{\mcitedefaultmidpunct}
{\mcitedefaultendpunct}{\mcitedefaultseppunct}\relax
\EndOfBibitem
\bibitem[Hu \latin{et~al.}(2019)Hu, Zheng, Ma, Lu, Zhang, Zhang, Xia, Hao, He,
  Chen, \latin{et~al.} others]{Hu2019}
Hu,~Y.; Zheng,~X.; Ma,~G.; Lu,~H.; Zhang,~L.; Zhang,~C.; Xia,~Y.; Hao,~Y.;
  He,~L.; Chen,~J.; others {Giant Negative Thermal Expansion in
  Antiferromagnetic CrAs-Based Compounds}. \emph{Phys. Rev. Appl.}
  \textbf{2019}, \emph{12}, 034027\relax
\mciteBstWouldAddEndPuncttrue
\mciteSetBstMidEndSepPunct{\mcitedefaultmidpunct}
{\mcitedefaultendpunct}{\mcitedefaultseppunct}\relax
\EndOfBibitem
\bibitem[Song \latin{et~al.}(2018)Song, Qiao, Huang, Wang, Liu, Li, Chen, and
  Xing]{Song2018}
Song,~Y.; Qiao,~Y.; Huang,~Q.; Wang,~C.; Liu,~X.; Li,~Q.; Chen,~J.; Xing,~X.
  Opposite thermal expansion in isostructural noncollinear antiferromagnetic
  compounds of Mn3A (A= Ge and Sn). \emph{Chem. Mater.} \textbf{2018},
  \emph{30}, 6236--6241\relax
\mciteBstWouldAddEndPuncttrue
\mciteSetBstMidEndSepPunct{\mcitedefaultmidpunct}
{\mcitedefaultendpunct}{\mcitedefaultseppunct}\relax
\EndOfBibitem
\bibitem[Nevidomskyy and Coleman(2009)Nevidomskyy, and
  Coleman]{Nevidomskyy-2009}
Nevidomskyy,~A.~H.; Coleman,~P. {Kondo Resonance Narrowing in $d$- and
  $f$-Electron Systems}. \emph{Phys. Rev. Lett.} \textbf{2009}, \emph{103},
  147205\relax
\mciteBstWouldAddEndPuncttrue
\mciteSetBstMidEndSepPunct{\mcitedefaultmidpunct}
{\mcitedefaultendpunct}{\mcitedefaultseppunct}\relax
\EndOfBibitem
\bibitem[Yin \latin{et~al.}(2011)Yin, Haule, and Kotliar]{Yin-2011}
Yin,~Z.; Haule,~K.; Kotliar,~G. {Kinetic frustration and the nature of the
  magnetic and paramagnetic states in iron pnictides and iron chalcogenides}.
  \emph{Nat. Mater.} \textbf{2011}, \emph{10}, 932--935\relax
\mciteBstWouldAddEndPuncttrue
\mciteSetBstMidEndSepPunct{\mcitedefaultmidpunct}
{\mcitedefaultendpunct}{\mcitedefaultseppunct}\relax
\EndOfBibitem
\end{mcitethebibliography}

\end{document}